\documentclass[12pt]{article}
\usepackage[dvips]{graphics}

\begin{document}

\medskip
\begin{center}

\Large{Weak nuclear forces cause \\
the strong nuclear force}

\vspace{0.5cm}

\large{E.L. Koschmieder}

\bigskip

\small{Center for Statistical Mechanics\\The University of Texas at 
Austin, Austin TX
78712.  USA\\
e-mail: koschmieder@mail.utexas.edu}

\end{center}
\bigskip

\noindent
\small
{We determine the strength of the weak nuclear force which holds the 
lattices of the elementary particles together. We also determine the strength
of the strong nuclear force which emanates from the sides of the nuclear 
lattices and binds one nucleon to another nucleon.
The strong force is the sum of the unsaturated weak forces at the 
surface of the lattices. The strong force is then about 10$^6$ times 
stronger than the weak force between two lattice points.}


\normalsize

\section*{Introduction}

After have explained the masses of the electron and muon and of the 
stable mesons and baryons with cubic lattices consisting of either
photons or of  neutrinos in [1],
we can now determine the strength of the weak and the strength of the 
strong nuclear forces. Both are 70 years old puzzles.
We have assumed in [1] that the cubic lattices of the elementary particles
are held together by a weak force which acts from each lattice point to its
nearest neighbors, analogous to the ionic lattices of salt crystals. It 
is not necessary to specify this weak force for the calculation of the rest mass 
of the particles. We will, in the following, use lattice theory to determine the 
strength of the weak force which holds the lattices of the elementary particles together. 
We will then show that the strong force between  two elementary particles 
is nothing but the sum of the unsaturated weak forces emanating from the lattice 
points at the surface of the lattice. Consequently the strong force between two 
elementary particles is on the order of $10^6$ times stronger than the weak 
force in the interior of the particles. The strong force is caused by the weak force.

\section{The force in the interior of the particles}

In order to determine the force in the interior of the cubic lattices with  
which we have explained the particle masses we will,
as we have done before in [2], use a classical paper by Born and 
Land\'{e}  [3], (B\&L), dealing with the potential and compressibility 
of regular crystals. It is essential to realize that,
\begin{itemize}
\item
 \emph{for the existence of a cubic lattice it is necessary that the force 
between the lattice points has an attractive part and a repulsive part}.
\end{itemize}
\begin{itemize}
\item 
{Otherwise the lattice would not be stable and collapse}.
\end{itemize}
For the ionic crystals considered by B\&L the Coulomb force between the ions
 is the attractive force, whereas the repulsive force originates from the electron 
clouds surrounding the ions. When the electron clouds of the ions approach
each other during the lattice oscillations they repel each other. The magnitude 
of the repulsive force is not known per se and has to be determined from the
properties of the crystal.

We will follow exactly the procedure in B\&L in order to see whether their 
theory is also applicable to a cubic lattice made of neutrinos. In this case the
Coulomb force is, of course, irrelevant. As B\&L do, we say that the potential 
 of a cell of the neutrino lattice has an attractive part $-$ a/$\delta$ and 
a repulsive part + b/$\delta^\mathrm{n}$ with the unknown exponent n. $\delta$ is the 
distance in the direction between two neutrinos of the same type, either
muon neutrinos and anti-muon neutrinos or electron neutrinos and 
anti-electron neutrinos.

   The potential of a cell in an ionic  cubic  lattice is of the form
\begin{equation}
\phi = -\,\mathrm{a}/\delta + \mathrm{b}/\delta^\mathrm{n}\,, \end{equation}
Eq.(1) of B\&L. The constant b is eliminated with the equilibrium condition 
d$\phi$/d$\delta$ = 0. Consequently 
\begin{equation} \mathrm{b} = \frac{\mathrm{a}}{\mathrm{n}}
\,\delta_0^{\mathrm{n} -1}\,, \end{equation}
where $\delta_0$ is the lattice constant.
The unknown exponent n of $\delta$ in Eq.(1) was determined by B\&L with
the help of the compression modulus $\kappa$ which is known for ionic 
crystals. $\kappa$ is defined by 
\begin{equation} \kappa = -\,1/V\cdot dV/dp\,. \end{equation}
The compression modulus  of an ionic lattice is given by  
\begin{equation} \kappa = 9\,\delta_0^4/\mathrm{a}\,(\mathrm{n} -1)\,, 
\end{equation}
Eq.(4) in B\&L.
The interaction constant a of the Coulomb force in cubic ionic
crystals resulting from the contributions of all ions on a single ion is given 
by Eq.(5) of B\&L, it is 
\begin{equation}\mathrm{a} = 13.94\,\mathrm{e}^2 = 3.2161\cdot10^{-18}\,\mathrm{erg\cdot cm}\,,\end{equation}
where e is the elementary electric charge. This equation is fundamental for
the theory of ionic lattices and is based on an earlier 
paper by Madelung [4]. Consequently we find that
\begin{equation} (\mathrm{n} -1) = 10.33\,\mathrm{r}_0^4/\mathrm{e}^2\kappa\,, 
\end{equation}
where r$_0$ = $\delta_0$/2 is the distance between a pair of neighboring
Na and Cl ions. For the alkali-halogenids, such as NaCl or KCl, B\&L found
that n $\approx$ 9. If n = 9 is used in Eq.(4) to determine theoretically the 
compression modulus $\kappa$, then the theoretical values of $\kappa$ 
agree, in a first approximation, with the experimental values of $\kappa$,
thus confirming the validity of the ansatz for the potential in Eq.(1).

\bigskip
We now apply Eq.(6) to the neutrino lattice of the elementary particles in
order to determine the potential in the interior of the particles. We must 
use for r$_0$  the distance between two neighboring neutrinos in the lattice,
which is equal to the range of the weak nuclear force. The range of the weak
force is 1$\cdot10^{-16}$ cm, according to p.25 of Perkins [5], and so we have
\begin{equation} \mathrm{r}_0 = 1\cdot 10^{-16}\,\mathrm{cm}\,. \end{equation}
We have used this value of r$_0$ throughout our 
explanation of the masses of the elementary particles in [1], though r$_0$ 
was then designated by the symbol \emph{a}. We must, furthermore, 
replace e$^2$ by the interaction constant g$_w^2$ of the weak force which
holds the nuclear lattice together. According to Perkins [5] 
\begin{equation} \mathrm{g}_w^2 = 4\pi\hbar\,\mathrm{c}\cdot 1.05\cdot10^{-2}\,
(\mathrm{M}_W/\mathrm{M}_p)^2\,,  \end{equation}
where M$_W$ is the mass of the W boson, M$_W$ = 80.403\,GeV/c$^2$,
and M$_p$ is the mass of the proton, M$_p$ = 0.938\,272\,GeV/c$^2$.
That means that 
\begin{equation} \mathrm{g}_w^2 = 2.9976\cdot 10^{-17}
\,\mathrm{erg\cdot cm}. \end{equation}

   We must also use the compression modulus $\kappa$ of the nucleon.
The value of the compression modulus of the nucleon  has been determined
theoretically by Bhaduri et al. [6], following earlier theoretical and experimental
investigations of the compression moduli of nuclei. Bhaduri et al. found that the
compression modulus K$_A$ of the nucleon ranges from 900 to 1200\,MeV,
or is 940\,MeV  or  900\,MeV for particular sets of parameters. We determine
$\kappa$ of the nucleon with
\begin{equation} \kappa = 9/\rho_\#\mathrm{K}_\mathrm{nm}\,,
\end{equation}
from Eq.(1) in [6]. Bhaduri et al. write that ``the compression modulus 
K$_\mathrm{nm}$ of nuclear matter is calculated by considering the 
nucleons as point particles", which they are not. Other assumptions are also 
sometimes made such as infinite nuclear matter, periodic boundary conditions, 
etc. Recent theoretical studies of the compressibility of ``nuclear matter" [7,8,9]
place the compressibility K$_\mathrm{{nm}}$ at values from between 
250 to 270 MeV, not much different from what is was twenty years earlier
in [6]. Considering the uncertainty of K$_\mathrm{nm}$ it seems to be 
justified to set, in the case of the nucleon, K$_\mathrm{nm}$ = 
K$_\mathrm{A}$, where K$_\mathrm{A}$ is defined as the compression 
modulus for a finite system with A nucleons. It then follows from Eq.(10)
with the number density $\rho_\#$ being the number density per fm$^3$,
and with the radius of the nucleon R$_0$ = 0.88\,$\cdot$\,10$^{-13}$\,cm, 
and with 1\,MeV = 1.6022\,$\cdot$\,10$^{-6}$\,erg that the compression 
modulus of the nucleon is  
\begin{equation} \kappa_n = 1.603\cdot10^{-35}\,\,\mathrm{cm^2/dyn}\,,
\end{equation}
if we use for K$_\mathrm{A}(\mathrm{N}$) the value 1000\,MeV. We will 
keep in mind 
that $\kappa_n$ is not very accurate because K$_A$(N) is not very accurate.

If we insert (7), (9), and (11) into n $-$ 1 = 10.33\,r$_0^4$/e$^2\kappa$
(Eq.6) we find an equation for the exponent  n of the term 
b/$\delta^\mathrm{n}$  in the repulsive part of the potential in a nuclear lattice, 
\begin{itemize}
\item
\begin{equation} \mathrm{n} = 1 + 2.164\cdot10^{-12} = 1 + \epsilon\,.
\end{equation}
\end{itemize}

That means that
 
\bigskip
\emph{the potential $\phi$ in the interior of an elementary 
particle is given by}
\begin{equation} \phi = - \,\frac{\mathrm{a}}{\delta} + \frac{\mathrm{b}}
{\delta^{1 + \epsilon}} = \frac{\mathrm{a}}{\delta}\,[\frac{(\delta_0/\delta)\,^\epsilon}{\mathrm{n}} - 1\,]\,,
\end{equation}
which we can reduce with n $-$ 1 = $\epsilon$, neglecting a term multiplied 
by $\epsilon\,^2$ = O(10$^{-24}$), using also a = 13.94\,e$^2$ (Eq.5) 
and 1/n $\cong$ (1$-$ $\epsilon$), to
\begin{equation} \phi \cong -\,\frac{\mathrm{a}\,\epsilon}{\delta}\,
[1 - \mathrm{ln}(\delta_0/\delta)] \cong -\,\frac{13.94\,\mathrm{g}_w^2\,\epsilon}
{\delta} \, [1 - \mathrm{ln}(\delta_0/\delta)\,]\,. \end{equation}
In equilibrium the value of $\phi$ in the nuclear lattice is about 
g$_w^2\cdot\epsilon$/e$^2$ $\approx$ 2.7$\cdot$10$^{-10}$ times 
smaller than the corresponding electrostatic potential in an ionic lattice. A 
graph of the potential in Eq.(14) versus $\delta$ is shown in Fig.\,1.

\begin{figure}[h]
	\vspace{0.5cm}
	\hspace{2.5cm}
	\includegraphics{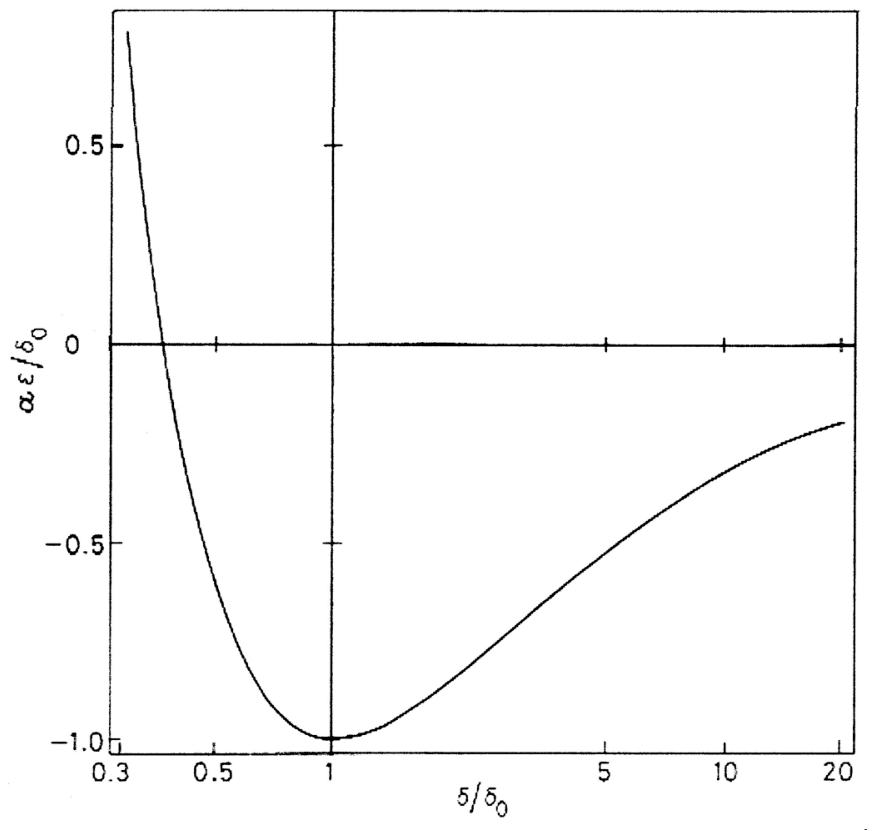}
	\vspace{-0.2cm}
	\begin{quote}
Fig.1: The potential $\phi$ of the weak force as a function of $\delta$.\\
\indent\hspace{1.1cm} After [2].
	\end{quote}
\end{figure}

\bigskip

 The minimum of the curve marks the equilibrium. The small value of 
$\epsilon$ $\approx$ 10$^{-12}$ means that small displacements 
$\delta_0/\delta$  $>$\,1 of the neutrinos from their equilibrium position, 
which carry the neutrinos into the domain of their neighboring neutrino, 
cause a very strong 
repulsive force between both neutrinos. From Eq.(14) follows with
F$_\mathrm{w}$ = d$\phi$/d$\delta$  and $\delta$ = 2r that 

\vspace{1cm} 
\hspace{1cm}\emph{the weak force in the interior of the nuclear lattice is}  

\begin{equation} \mathrm{F}_\mathrm{w} \cong -\,\frac {3.485\cdot \mathrm{g}_w^2\,\epsilon}
{\mathrm{r}^2}\cdot \mathrm{ln}(\frac{\delta_0}{\delta})\,. 
\qquad(\mathrm{dyn})\end{equation}
For all distances $\delta$ $>$ $\delta_0$ the force F$_\mathrm{w}$ is attractive, 
for all distances $\delta$ $<$ $\delta_0$ the force is repulsive.

   We have thus determined the potential of the weak force in the interior of 
the lattice of the elementary particles with lattice theory. Let us consider how 
this was done. Following exactly the procedure used by B\&L to determine 
the potential in the interior of an ionic crystal, we have determined the 
potential  in the interior of the lattice of an
elementary particle by using the parameters of the nuclear lattice. As in an
ionic lattice the potential in a nuclear lattice has an attractive and a repulsive 
part, as is necessary for the stability of the lattice. We have found that in
equilibrium in a nuclear lattice the absolute values of the attractive and 
repulsive terms of the potential are very nearly the same, because 
$\epsilon$ is on the order of 10$^{-12}$. One wonders, in view of the 
extraordinaryly small value of $\epsilon$, whether the potential in Eq.(14)
is not an academic result. Therefore it has to be shown that Eq.(14) is 
indeed relevant for elementary particles. We can do so by showing that
the mass of the muon neutrino m($\nu_\mu$) depends on n $-$ 1 = 
$\epsilon$ and yields a correct mass m($\nu_\mu$) if only n $-$ 1 is on 
the order of 10$^{-12}$.

\bigskip
   From the explanation of the $\pi^\pm$\,mesons in [1] follows that the 
mass of the muon neutrinos is
\begin{equation} \mathrm{m}(\nu_\mu) = \mathrm{m}(\bar{\nu}_\mu)
\cong 50\,\mathrm{milli\,eV/c^2}\,, \end{equation}
Eq.(40) in [1]. One can, on the other hand, determine the mass of the muon 
neutrinos from the group velocity of the neutrino oscillations in their nuclear
lattice, more precisely from the limitation of the group velocity to values 
c$_g$ $\cong$ c, where c is the velocity of light. It follows from Eq.(26) of [1] that
\begin{equation} \mathrm{m}(\nu_\mu) \cong \mathrm{r}_0^3\,\mathrm{c}_{44}\,
/ \mathrm{c}^2\,, \end{equation}
where c$_{44}$ is one of the elastic constants of continuum mechanics, which is
valid when r$_0 \rightarrow$ 0. In the case of a central force between the lattice 
points and with isotropy we have 3\,c$_{44}$ = c$_{11}$. The elastic constant 
c$_{11}$ has been determined theoretically with lattice theory by Born [10],
Eq.(457) therein. From his theory follows, as shown in Eq.(17) of [2], that
\begin{equation} \mathrm{c}_{11} = 
0.538\,\mathrm{g}_w^2(\mathrm{n} - 1)/\mathrm{r}_0^4 = 3.466\cdot10^{35}
\,\,\mathrm{dyn/cm^2}\,. \end{equation}
 The elastic constant c$_{11}$  must be inversely proportional to $\kappa$, 
because c$_{11}$ + 2c$_{12}$ = 3/$\kappa$ and with Eq.(11) and 
Eq.(17) we have c$_{11}$  = 5.556/$\kappa$.  From Eq.(17) follows with 
3\,c$_{44}$ = c$_{11}$ and Eq.(18) that
\begin{equation} \mathrm{m}(\nu_\mu) = 0.538\,\mathrm{g}_w^2
(\mathrm{n} -1)/3\mathrm{r}_0\mathrm{c}^2\,. \end{equation}
The mass of the muon neutrino depends, other than on known constants,
only on (n $-$ 1). Using Eqs.(7,9,12) we arrive at
\begin{equation}
\mathrm{m}(\nu_\mu) = 1.284\cdot 10^{-34} \mathrm{gr} = 72.1\,\mathrm{milli\,eV}/\mathrm{c}^2 = \mathrm{m}(\bar{\nu}_\mu)\,.
 \end{equation}
because the anti-muon neutrinos in the lattice move with the same  group 
velocity as  the muon neutrinos.  The value of m($\nu_\mu)$  in Eq.(20) 
differs by  50\%  from m($\nu_\mu)$  $ \cong$ 50\,milli-eV/c$^2$ given 
by Eq.(16).  We have, however,
to consider the large uncertainty of K$_A$. If we would use K$_A$(N) =
700\,MeV in the determination of (n $-$ 1) then we arrive at m($\nu_\mu$)
$\approx$ 50\,milli-eV/c$^2$, as it must be. Rather than by the value of
(n $-$ 1) the accuracy of the value of m($\nu_\mu$) is determined by the 
value of K$_A$(N). We have learned, however, that \emph{$\epsilon$ 
must be on the order of 10$^{-12}$ to end up with a correct value of 
the muon neutrino mass m($\nu_\mu$)}.

To summarize:  We have found that it follows from lattice theory that the 
value of
the exponent n = 1 + $\epsilon$ in the repulsive term of the potential in 
the nuclear lattice leads to an, in the first approximation, correct value of the
 mass of the muon neutrino, albeit that (n $-$ 1) is so extraordinarily small.
We have thus confirmed the validity of the potential $\phi = a/\delta$ + 
b/$\delta^\mathrm{n}$.
That means \emph{we have found the potential of 
the weak force which holds the nuclear lattice together}.
The astonishing fact that the mass of the muon neutrino can be derived from 
the group velocity of the oscillations in the particle lattices demonstrates the 
power of lattice theory for the explanation of the elementary particles.

\section{The strong force between two elementary particles} 

   Crucial for the understanding of the existence of a strong force between 
the sides of two cubic lattices is the observation that
\begin{itemize}
\item
 \emph{the sides of  two halves of a cubic lattice cleaved in vacuum 
exert a strong, attractive force on each other}.
\end{itemize}
 It is an automatic consequence of lattice theory that
\begin{itemize}
\item
\emph{the weak force, which holds the lattice together, 
is accompanied by a strong force which emanates from 
the sides of the lattice}.
\end{itemize} 
 This seems to contradict the simple 
observation that two salt crystals stacked upon each other can be 
separated without any effort. This is only so because the surface of 
a cubic crystal cleaved in air oxidizes immediately, and then the sides  
do not attract each other any longer. The origin of the force emanating 
from the sides of a cubic ionic lattice is the Coulomb force between
ions of opposite polarity, i.e.\,\,the force which
 holds the lattice together. The attractive force emanating from
the side of a crystal cleaved in vacuum has a macroscopic value. The 
existence of this force has tangible consequences in space technology.
The force between the sides of two cubic lattices was first studied 
by Born and Stern [11] (B\&S).

   If U$_{12}$ is the potential between two halves of a crystal with the 
equal surfaces A, or the work that is necessary to separate the two halves 
of a cleaved crystal, then the capillary constant $\sigma$ is given by
Eq.(2) of B\&S 
\begin{equation} \sigma = -\,\mathrm{U}_{12}/\mathrm{A}\,. \end{equation}
The capillary constant is, in the following, taken at zero degree absolute 
and against vacuum for the square outside area A of a cubic crystal. $\sigma$ 
has been explained by B\&S, but their formula cannot be 
used here because they use the value  n = 9 of the alkali-halegonids.
We will instead use Eq.(463) from Born [10] for the capillary constant 
$\sigma_{100}$ of the (100) front surface of an ionic cubic crystal
\begin{equation} \sigma_{100} = -\,\frac{\mathrm{e}^2}{\mathrm{r}_0^3}\,
\frac{s_0(1)}{2}\cdot[\,1 - \frac{1}{\mathrm{n}}\,\frac{s_0(\mathrm{n})}
{s_0(1)}\cdot\frac{S_0(1)}{S_0(\mathrm{n})}\,]\,. \end{equation}
The sums s$_0$(n) and S$_0$(n) originate from the contributions of the 
different lattice points to the repulsive part of the potential. The sign of the 
second term on the right hand side in Eq.(22) comes from the repulsive 
part of the potential in Eq.(1). For the capillary constant in a nuclear lattice 
we set e$^2$ =  g$_w^2$, n = 1 + $\epsilon$, (Eq.12), and  
s$_0$(1) = $-$\,0.0650 from [10] p.743. We find that 
s$_0$(n) $\cong$ s$_0$(1) since n = 1 + $\epsilon$ and 
$\epsilon \cong$ 10$^{-12}$.  Similarly we have 
S$_0$(n) $\cong$ S$_0$(1). Then we have
\begin{equation} \sigma_{100} \cong \frac{0.065}{2}\,\frac{\mathrm{g}_w^2}
{\mathrm{r}_0^3}\, \epsilon\,. \qquad(\mathrm{dyn/cm}) \end{equation}

The work required to separate one half of the lattice from the other  
half is according to Eq.(21) given by 
\begin{equation} \mathrm{U}_{12} \cong -\,\frac{0.065}{2}\,\, 
\frac{\mathrm{g}_w^2\,\epsilon}{\mathrm{r}_0^3}\cdot\mathrm{A}\,.
\end{equation}

\medskip    
\noindent
And it follows that the strong attractive force between the sides
of two nuclear lattices  is
\begin{equation} \mathrm{F_s} = \frac{\mathrm{dU}_{12}}{\mathrm{dr}} =
-\,\frac{\mathrm{d}\sigma}{\mathrm{dr}}\cdot\mathrm{A} = 
\frac{3\cdot0.065\,\mathrm{g}_w^2\,\epsilon}
{2\,\mathrm{r}^4}\cdot\mathrm{A}\,. \end{equation}

\medskip
\noindent
With N being the number of the lattice points in the cubic nuclear lattice, 

\begin{equation} \mathrm{N}  = 2.854\cdot10^9\,, \end{equation}

\medskip
\noindent
from Eq.(15) in [1], and with A = ($\sqrt[3]{\mathrm{N}}\,\cdot\,$r$_0)^2$ 
follows that 

\medskip
\emph{the force emanating from the front surface of a cubic nuclear lattice,
\indent 
the strong force, is} 
\begin{equation} \mathrm{F_s} = \frac{0.0975\,\,\mathrm{g}_w^2\,\,\epsilon}
{\mathrm{r}^4}\cdot(\sqrt[3]{\mathrm{N}}\cdot\mathrm{r}_0)^2\,. 
 \qquad (\mathrm{dyn}) \end{equation}

\medskip
\noindent
The strong force depends first of all on the \emph{weak interaction constant}
g$_w^2$, and secondly on the number of lattice points on the side of 
the lattice, via ($\sqrt[3]{\mathrm{N}})^2\,$. The strong force is also inversely 
proportional to the \emph{fourth} power of the distance between the particles.
The same force goes out from each side of the lattice, the entire force 
which goes out from the surface of the lattice is consequently six times 
as much as given by Eq.(27). At  r = 1\,$\cdot\,$10$^{-16}$\,cm away  
from one side of the surface of this lattice the strong force is
\begin{equation} \mathrm{F_s} = 1.27\cdot10^9\,\,\mathrm{dyn}\,.
\end{equation}

\newpage
The ratio of the strong force F$_\mathrm{s}$ emanating from the surface 
of a cubic nuclear lattice to the weak force F$_\mathrm{w}$ between the 
neutrinos in the same lattice is
\begin{equation} \mathrm{F_s}/\mathrm{F_w} = 6\cdot 
0.028\cdot (\sqrt[3]{\mathrm{N}}\,\mathrm{r}_0)^2/\mathrm{r^2\,ln}(\delta_0/\delta) = 
0.338\cdot10^6\,\mathrm{r}_0^2/\mathrm{r^2\,ln}(\delta_0/\delta)\,.\end{equation}
The prime factor in the ratio of the strong and weak forces is the number of the lattice points at a side of the lattice $(\sqrt[3]{\mathrm{N}})^2$ = 2.012$\cdot10^6$.

   To summarize: According to lattice theory 
\begin{itemize}
\item
\emph{the existence of the 
strong nuclear force between two elementary particles is an automatic
consequence of our explanation of the masses of the elementary
particles with cubic nuclear lattices,} 
\end{itemize}
held together by the weak nuclear
force. The strong force between the sides of two nuclear 
lattices is a multiple of the unsaturated part of the weak force at each
lattice point at the surface of the lattice. The strong and the weak force
 have the same origin, the weak attraction of the lattice points to its 
nearest neighbors. The lattices we have used for the explanation of 
the masses of the particles consist of photons or neutrinos. That means: 
\begin{center}
\emph{We do not use hypothetical particles.}  
\end{center}

\bigskip
   In qualitative terms: Only about one sixth of the force going out from each 
lattice point at the surface of the lattice is unsaturated and directed outward, the
other 5/6 of the force between the lattice points are saturated by the five 
neighboring lattice points. That means, in this qualitative approximation,
that the sum of the forces going out from the surface of the lattice is  the
number of lattice points at the surface times 1/6 of 
the force between each lattice point. The strong force going 
out from the surface should therefore be equal to the number 
of the lattice points at the surface, or equal to ($\sqrt[3]{\mathrm{N }})^2$ = 
2.012$\,\cdot\,10^6$, times \,$\approx$\,1/6 of the weak force between 
the lattice points. When we compare the ratio F$_\mathrm{s}$/F$_\mathrm{w}
$ = 0.336\,$\cdot\,10^6$ from 
Eq.(29) with an F$_\mathrm{s}$ = 
($\sqrt[3]{\mathrm{N}})^2$/6\,$\cdot$\,F$_\mathrm{w}$ = 0.335\,$\cdot$\,10$^6$\,F$_\mathrm{w}$ we find that the assumption that 1/6 
of F$_\mathrm{w}$ at the surface is unsaturated is a reasonable approximation. 
The factor ($\sqrt[3]{\mathrm{N}})^2$ in Eq.(29) tells that the strong force 
emanating from a cubic  lattice
is the sum of the unsaturated weak forces of the lattice points at the sides of 
the lattice. A precise experimental value of the ratio of the strong nuclear force  
to the weak nuclear force does not seem to be known. It is, however, usually 
said that the ratio of the strong interaction constant $\alpha_s$ to the weak 
interaction constant $\alpha_w$ is $\alpha_s/\alpha_w \approx 10^6$.

\section*{Conclusions}

We have found long sought after answers to the questions what is the weak 
nuclear force and the strong nuclear force, and why is the strong force so much
stronger than the weak force\,? To arrive at our answers we have used  
results of lattice theory, according to which an ionic cubic lattice 
presents two kinds of forces, one weak force by which a lattice point  
is bound to its nearest neighbors,
and a second much larger force which emanates from the sides of
a lattice cleaved in vacuum. If the elementary particles consist of a nuclear
lattice then a strong force necessarily emanates from the sides of the lattice.
Inserting into the results of the theory of ionic cubic
lattices the relevant parameters of the nuclear lattice, in particular the weak
interaction constant g$_w^2$, and the nuclear lattice distance r$_0$ which 
follows from the range of the weak nuclear force, we arrive at formulas for  
the strength of the weak and the strength of the strong nuclear forces and 
find that the strong force is on the order of 10$^6$  times
stronger than the weak force. The strong force is nothing but the sum 
of the large number of unsaturated weak forces at the surface of   
the lattice. In order to understand the origin of the strong nuclear force 
one has to understand the structure of the elementary particles, which 
we have explained in [1] with nuclear lattices. With this concept we can
explain the masses of the particles and the spin of the 
particles. We have now also understood the strength of the weak force
which holds the nuclear lattice together and the cause of the strong force
between two nuclear lattices.

\section*{References}                   
[1] Koschmieder, E.L. http://arXiv.org:physics/0602037. (2006)

\smallskip
\noindent
[2] Koschmieder, E.L. Nuovo Cim. {\bfseries101A},1017. (1989)

\smallskip
\noindent
[3] Born, M. and Land\'{e}, A. Verh.Dtsch.Phys.Ges. {\bfseries20},210. (1918)

\smallskip
\noindent
[4] Madelung, E. Phys.Z. {\bfseries19},524. (1918)

\smallskip
\noindent
[5] Perkins, D.H. \emph{Introduction to High-Energy Physics}\\
\indent Addison-Wesley. (1982)

\smallskip
\noindent
[6] Badhuri, R.K. et al. Phys.Lett.B {\bfseries136},189. (1984)

\smallskip
\noindent
[7] M\"{u}ller, H-M. et al. http://arXiv.org:nucl-th/9910038. (1999)

\smallskip
\noindent
[8] Vretenar, D. et al. http://arXiv.org:nucl-th/0302070. (2003)

\smallskip
\noindent
[9] Dexheimer, V.A. et al. http://arXiv.org:0708.0131. (2007)

\smallskip
\noindent
[10] Born, M. \emph{Atomtheorie des festen Zustandes}\\
\indent Teubner. (1923)

\smallskip
\noindent 
[11] Born, M. and Stern, O. Sitzungsber.Preuss.Akad.Wiss. {\bfseries33},901. (1919)

\end{document}